\documentclass{ws-p8-50x6-00}

\begin{document}

\title{Traces of nonextensivity in particle physics due to
fluctuations}

\author{G.Wilk}

\address{The Andrzej So\l tan Institute for Nuclear Studies; 
Ho\.za 69; 00-689 Warsaw, Poland\\E-mail: wilk@fuw.edu.pl}

\author{Z.W\L odarczyk}

\address{Institute of Physics, \'Swi\c{e}tokrzyska Academy;
Konopnickiej 15; 25-405 Kielce, Poland\\
E-mail: wlod@pu.kielce.pl}


\maketitle

\abstracts{We present a short review of traces of nonextensivity in
particle physics due to fluctuations. 
}

\section{Introduction: connection of fluctuations and nonextensivity}

Both the notion of fluctuations and that of nonextensivity are
nowdays widely known, albeit mostly in the fields of research only
indirectly connected with particle physics. Both turns out to be very
fruitful and interesting and this is clearly demonstrated by all
other lectures given at this workshop (see especially \cite{B,T}). 

This lecture will be devoted to the case in which evident 
nonextensivity of some expressions originate in intrinsic
fluctuations in the system under consideration (the origin of which
is usually not yet fully understood)\footnote{Our encounter with this
problem is presented in works
\cite{CR,WW,DENTON,UWW,OFT,UWWF,NEXPDEC,RWW,NEXT,CERN}.}. The best 
introduction to this problem is provided by the observation that in
some cosmic ray data (like depth distribution of starting points of
cascades in Pamir lead chamber \cite{WWCR}) one observes clear
deviations from the naively expected exponential distributions of 
some variables which are evidently better fitted by the power-like
formulas: 
\begin{equation}
\frac{dN}{dh}\, =\, {\rm const}\cdot \exp \left(\, -\,
\frac{h}{\lambda}\, \right) \Rightarrow
{\rm const}\cdot \left[1\, -\,
      (1-q)\frac{h}{\lambda}\right]^{\frac{1}{1-q}}. \label{eq:EXP}
\end{equation}
Here $N$ denotes the number of counts at depth $h$ (cf. \cite{WWCR}
for details). Whereas in \cite{WWCR} we have proposed as explanation
a possible fluctuations of the mean free path $\lambda$ in eq.
(\ref{eq:EXP}) characterised by relative variance $\omega\, =\,
\frac{\left(\langle \sigma^2\rangle - \langle \sigma \rangle
^2\right)}{\langle \sigma \rangle ^2}\, \geq\, 0.2 $,
in \cite{CR} the same data were fitted by power-like (L\'evy type)
formula as above keeping $\lambda$ fixed and setting $q=1.3$. In this
way we have learned about Tsallis statistics and Tsallis nonextensive
entropy and distributions\footnote{See Tsallis lecture \cite{T} and
references therein (cf. also \cite{DENTON,NEXT}) for detailed
necessary information concerning Tsallis statistics and the
non-extensivity.}. By closer inspection of the above example we have
been able to propose a new physical meaning of the nonextensivity
parameter $q$, as a measure of intrinsic fluctuations in the system
\cite{WW,DENTON}. Fluctuations are therefore proposed as a new source
of nonextensivity which should be added \cite{Becka} to the
previously known and listed in literature sources of the
nonextensivity (like long-range correlations, memory effects or 
fractal-like structure of the corresponding phases space \cite{T}) .

To demonstrate this conjecture let us notice that for $q>1$ case,
where $\varepsilon \in (0, \infty)$, one can write a kind of Mellin
transform (here $\alpha = \frac{1}{q-1}$) \cite{DENTON}:
\begin{eqnarray}
\left( 1\, +\, \frac{\varepsilon}{\lambda_0}\,
\frac{1}{\alpha}\right)^{-a}\, &=&\, 
 \frac{1}{\Gamma(\alpha)}\, \int^{\infty}_0\, d\eta\, 
      {\eta}^{\alpha - 1}\, \exp\left[ - \eta\, 
      \left(1\, +\, \frac{\varepsilon}{\lambda_0}\,
       \frac{1}{\alpha}\right)\right]\nonumber\\
&=&\, \int^{\infty}_0\, \exp\left( - \frac{\varepsilon}{\lambda}\right)\,
f_{q>1}\left(\frac{1}{\lambda}\right)\,
d\left(\frac{1}{\lambda}\right)\, ,
\label{eq:DEF+} 
\end{eqnarray}
where $f_{q>1}(1/\lambda)$ is given by the following gamma distribution:
\begin{equation}
f_{q>1}\left(\frac{1}{\lambda}\right)\, =\,
f_{q>1}\left(\frac{1}{\lambda},\frac{1}{\lambda_0}\right)\, =\,
\frac{\mu}{\Gamma(\alpha)}\,
\left(\frac{\mu}{\lambda}\right)^{\alpha-1}\, \exp\left(
- \frac{\mu}{\lambda}\right) \label{eq:F}
\end{equation}
with $\mu = \alpha \lambda_0$ and with mean value and variation in the form:
\begin{equation}
\left\langle \frac{1}{\lambda}\right\rangle \, =\,
 \frac{1}{\lambda_0} \qquad {\rm and}\qquad
\left\langle \left(\frac{1}{\lambda}\right)^2\right\rangle\, -\, 
\left\langle\frac{1}{\lambda}\right\rangle^2\, =\, 
\frac{1}{\alpha\, \lambda_0^2} . \label{eq:MEANVAR}
\end{equation}
For the $q<1$ case $\varepsilon$ is limited to $\varepsilon \in
[0,\lambda_0/(1-q)]$. Proceeding in the same way as before (but with
$\alpha' = - \alpha = \frac{1}{1-q}$) one gets:
\begin{eqnarray}
\left[1\, -\, \frac{\varepsilon}{\alpha' \lambda_0}\right]^{\alpha'}\, 
&=&\,
\frac{1}{\Gamma(\alpha')}\, \int^{\infty}_0\, d\eta\, 
    \eta^{\alpha' - 1}\, \exp\left[ - \eta\, \left(1\, +\, 
    \frac{\varepsilon}{\alpha' \lambda_0 -
\varepsilon}\right)\right]\nonumber\\ 
&=&\, \int^{\infty}_0\, \exp\left( - \frac{\varepsilon}{\lambda}\right)\,
f_{q<1}\left(\frac{1}{\lambda}\right)\,
d\left(\frac{1}{\lambda}\right)\, ,
\label{eq:DEF-} 
\end{eqnarray}
where $f_{q<1}(1/\lambda)$ is given by the same gamma distribution
as in (\ref{eq:F}) but this time with $\alpha \rightarrow \alpha'$
and $\mu = \mu(\varepsilon) = \alpha' \lambda_0 - \varepsilon$.
Contrary to the $q>1$ case, this time the fluctuations depend on the
value of the variable in question, i.e., the mean value and variance
are now both $\varepsilon$-dependent:  
\begin{equation}
\left\langle \frac{1}{\lambda}\right\rangle\, =\, \frac{1}{\lambda_0 -
\frac{\varepsilon}{\alpha'}}\qquad {\rm and}\qquad \left\langle
\left(\frac{1}{\lambda}\right)^2\right\rangle\, -\,
\left\langle\frac{1}{\lambda}\right\rangle^2\, =\, \frac{1}{\alpha'}\cdot
\frac{1}{\left(\lambda_0 - \frac{\varepsilon}{\alpha'}\right)^2} .
\label{eq:MV}
\end{equation}
However, in both cases the relative variances,
\begin{equation}
\omega\, =\, \frac{\left\langle\left(\frac{1}{\lambda}\right)^2\right\rangle\, 
 -\, \left\langle\frac{1}{\lambda}\right\rangle^2}
 {\left\langle \frac{1}{\lambda}\right\rangle^2}\, =\,
 \left\{ \begin{array}{c}
\frac{1}{\alpha}\\
\\ \frac{1}{\alpha'} \end{array} \right\} 
\, =\, \left\{ \begin{array}{c} 
q - 1\\ 
\\ 1 - q \end{array} \right\}\, , \label{eq:PROOF}
\end{equation}
remain $\varepsilon$-independent and depend only on the parameter
$q$ \footnote{Notice that, with increasing $\alpha$ or $\alpha'$
(i.e., for $q\rightarrow 1$) both variances (\ref{eq:PROOF}) 
decrease and asymptotically gamma distributions (\ref{eq:F}) becomes
a delta function, $f(1/\lambda)=\delta (\lambda - \lambda_0)$.}.
It means therefore that \cite{WW,DENTON} (at least for the
fluctuations distributed according to gamma distribution)
\begin{equation}
L = \exp\left(-\frac{x}{\lambda_0}\right)~~ \Rightarrow~~
L_q = \exp_q\left(-\frac{x}{\lambda_0}\right) =
\left\langle \exp\left(-\frac{x}{\lambda}\right)\right\rangle ,
\label{eq:Nonext}
\end{equation}
with $q = 1 \, \pm \, \omega $ for $q>1~(+)$ and $q<1~(-)$, i.e.,
there is connection between the measure of fluctuations $\omega$ and
the measure of nonextensivity $q$ (it has been confirmed recently in
\cite{Becka}).

\section{Where are the fluctuations coming from?}

\subsection{Generalities}

The proposed interpretation of the parameter $q$ leads immediately to
the following question: {\it why and under what circumstances it is
the gamma distribution that describes fluctuations of the parameter
$\lambda$?}  To address this question let us write the usual Langevin
equation for the stochastic variable $\lambda$ \cite{WW,DENTON}:
\begin{equation}
\frac{d\lambda}{dt}\, +\, \left[\frac{1}{\tau}\, +\, \xi(t)\right]\,
\lambda\, =\, \phi\, =\, {\rm const}\, >\, 0 . \label{eq:LE}
\end{equation}
with damping constant $\tau$ and with source term $\phi$, different
for the two cases considered, namely:  
\begin{equation}
\phi = \phi_{q<1}\, =\, \frac{1}{\tau}\left(\lambda_0 -
                 \frac{\varepsilon}{\alpha'}\right) 
\qquad {\rm whereas}\qquad \phi = \phi_{q>1} = \frac{\lambda_0}{\tau} .
             \label{eq:FIFI}
\end{equation}
For the usual stochastic processes defined by the {\it white gaussian
noise} form of $\xi(t)$ \footnote{It means that ensemble mean
$\langle \xi(t) \rangle\, =\, 0 $ and correlator (for sufficiently
fast changes) $\langle \xi(t)\, \xi(t + \Delta t) \rangle\, =\, 2\,
D\, \delta(\Delta t)$. Constants $\tau$ and $D$ define, respectively,
the mean time for changes and their variance by means of the
following conditions: $\langle \lambda(t)\rangle\, =\, \lambda_0\,
\exp\left( - \frac{t}{\tau} \right)$ and $\langle
\lambda^2(t=\infty)\rangle\, =\, \frac{1}{2}\, D\, \tau$.
Thermodynamical equilibrium is assumed here (i.e., $t >> \tau$, in
which case the influence of the initial condition vanishes and the 
mean squared of $\lambda$ has value corresponding to
the state of equilibrium).} one obtains the following Fokker-Plank
equation for the distribution function of the variable $\lambda$  
\begin{equation}
\frac{df(\lambda)}{dt}\, =\, -\, \frac{\partial}{\partial \lambda}K_1\,
f(\lambda)\, +\, \frac{1}{2}\, \frac{\partial^2}{\partial \lambda^2}K_2\,
f(\lambda) , \label{eq:FPE}
\end{equation}
where the intensity coefficients $K_{1,2}$ are defined by
eq.(\ref{eq:LE}) and are equal to \cite{WW,DENTON}: 
\begin{equation}
K_1(\lambda)\, =\, \phi\, -\, \frac{\lambda}{\tau}\, +\, D\, \lambda
\qquad {\rm and}\qquad 
K_2(\lambda)\, =\, 2\, D\, \lambda^2 . \label{eq:KK}
\end{equation}
From it we get the following  expression for the distribution
function of the variable $\lambda$:
\begin{equation}
f(\lambda)\, =\, \frac{c}{K_2(\lambda)}\, \exp\left[\, 2\,
\int^{\lambda}_0 d\lambda'\, \frac{K_1(\lambda')}{K_2(\lambda')}\, \right]
\label{eq:EF} 
\end{equation}
which is, indeed, a gamma distribution (\ref{eq:F}) in variable
$1/\lambda$, with the constant $c$ defined by the normalization
condition, $\int^{\infty}_0 d(1/\lambda) f(1/\lambda) = 1$, and
depending on two parameters: $\mu(\varepsilon)\, =\,
\frac{\phi_q(\varepsilon)}{D}$ and $\alpha_q\, =\, \frac{1}{\tau\,
D}$ with $\phi_q = \phi_{q>1,q<1}$ and $\alpha_q = (\alpha, \alpha')$
for, respectively, $q>1$ and $q<1$. This means that we have obtained
eqs.(\ref{eq:PROOF}) with $\omega = \frac{1}{\tau D}$ and, therefore,
the parameter of nonextensivity $q$ is given by the parameter $D$ and
by the damping constant $\tau$ describing the {\it white noise}.

\subsection{Temperature fluctuations}

The above discussion rests on the stochastic equation (\ref{eq:LE}).
Therefore the previously asked question is not yet fully answered but
can be refrazed in the following way: {\it can one point the possible
physical situation where such fluctuations could be present in the
realm of particle physics?} Our proposition to answer it is to
identify $\lambda = T$, i.e., to concentrate on the possibility of
fluctuations of temperature widely discussed in the literature in
different contexts \cite{TEMP}.  In all cases of interest to us
temperature $T$ is variable encountered in statistical descriptions
of multiparticle collision processes \cite{QCD}. Our reasoning goes
then as follows: Suppose that we have a thermodynamic system, in a
small (mentally separated) part of which the temperature fluctuates
with $\Delta T \sim T$. Let $\lambda(t)$ describe stochastic changes
of the temperature in time. If the mean temperature of the system is
$\langle T\rangle = T_0$ then, as result of fluctuations in some
small selected region, the actual temperature equals $T' = T_0 - \tau
\xi(t) T$ and the inevitable exchange of heat between this selected
region and the rest of the system leads to the process of
equilibration of the temperature which is described by the following
equation of the type of eq. (\ref{eq:LE}) \cite{LLH} : 
\begin{equation}
\frac{\partial T}{\partial t}\, -\, \frac{1}{\tau}
            \, (T'\, -\, T)\, +\, \Omega_q = 0 \label{eq:HC}
\end{equation}
(here $\Omega_{q<1} = \frac{\varepsilon}{\tau \alpha'}$ and
$\Omega_{q>1} = 0$). In this way we have demonstrated that eq.
(\ref{eq:LE}) can, indeed, be used  in statistical models of
multiparticle production as far as they allow the $q$-statistic as
their basis. 

The $q<1$ case requires a few additional words of explanation. Notice
the presence of the internal heat source, which is dissipating (or
transfering) energy from the region where (due to fluctuations) the
temperature $T$ is higher to colder ones. It could be any kind of 
convection-type flow of energy; for example, it could be connected
with emission of particles from that region. The heat release given
by $\varepsilon/(\tau\alpha')$ depends on $\varepsilon$ (but it is
only a part of $\varepsilon$ that is released). In the case of such
energy release (connected with emission of particles) there is
additional cooling of the whole system. If this process is
sufficiently fast, it could happen that there is no way to reach a
stationary distribution of temperature (because the transfer of  heat
from the outside can be not sufficient for the development of the
state of equilibrium). On the other hand (albeit this is not our case
here) in the reverse process we could face the "heat explosion"
situation (which could happen if the velocity of the exothermic
burning reaction grows sufficiently fast; in this case because of
nonexistence of stationary distribution we have fast nonstationary
heating of the substance and acceleration of the respective
reaction)\footnote{It should be noticed that in the case of $q<1$
temperature does not reach stationary state because, cf. Eq.
(\ref{eq:MV}), $\langle 1/T \rangle\, =\, 1/(T_0 -
\varepsilon/\alpha')$, whereas for $q>1$ we had $<1/T> = 1/T_0$. As a
consequence the corresponding L\'evy distributions are defined only
for $\varepsilon \in(0, T_0\, \alpha'$) because for $\varepsilon
\rightarrow T_0\alpha'$, $<T>\rightarrow 0$. Such asymptotic (i.e.,
for $t/\tau \rightarrow \infty$) cooling of the system ($T\rightarrow
0$) can be also deduced form Eq. (\ref{eq:HC}) for $\varepsilon
\rightarrow T_0\alpha'$.}.

The most interesting example of the possible existence of such
fluctuations is the trace of power like behaviour of the
transverse momentum distribution in multiparticle production
processes encountered in heavy ion collisions \cite{ALQ,UWW}. Such
collisions are of special interest because they are the only place
where new state of matter, the Quark Gluon Plasma, can be produced
\cite{QCD}. Transverse momentum distributions are believed to provide
information on the temperature $T$ of reaction, which is given by the
inverse slope of $dN/dp_T$, if it is exponential one. If it is not,
question arises what we are really measuring. One explanation is the
possible flow of the matter, the other, which we shall follow here,
is the nonextensivity (or rather fluctuations leading to it).
Namely, as was discussed in detail in \cite{ALQ} the extreme
conditions of high density and temperature occuring in
ultrarelativistic heavy ion collisions can invalidate the usual BG
approach and lead to $q>1$, i.e., to
\begin{equation}
\frac{dN(p_T)}{dp_T} = {\rm const}\cdot \left[ 1\, - (1 - q)
\frac{\sqrt{m^2 + p^2_T}}{T}\right]^{\frac{1}{1-q}}. \label{eq:Pt}
\end{equation}
Here $m$ is the mass of produced particle and $T$ is, for the $q=1$
case, the {\it temperature} of the hadronic system produced. Although
very small ($|q-1| \sim 0.015$) this deviation, if interpreted
according to eq. (\ref{eq:Nonext})), leads to quite large relative
fluctuations of temperature existing in the nuclear collisions,
$\Delta T/T \simeq 0.12$. It is important to stress that these are
fluctuations existing in small parts of hadronic system in respect to
the whole system rather than of the event-by-event type for which,
$\Delta T/T = 0.06/\sqrt{N} \rightarrow 0$ for large $N$ (cf.
\cite{OFT} for relevant references). Such fluctuations are
potentially very interesting because they provide direct measure of
the total heat capacity $C$ of the system,
\begin{equation}
\frac{\sigma^2(\beta)}{\langle \beta\rangle ^2}\, =\, \frac{1}{C}\,
=\, \omega\, =\, q - 1 \, ,\label{eq:C}
\end{equation}
($\beta =\frac{1}{T}$) in terms of $\omega = q - 1$. Therefore,
measuring {\it both} the temperature of reaction $T$ and (via
nonextensivity $q\neq 1$) its total heat capacity $C$, one can not
only check whether an approximate thermodynamics state is formed in a
single collision but also what are its theromdynamical properties
(especially in what concerns the existence and type of the possible
phase transitions \cite{OFT}).

\begin{figure}

  \begin{minipage}[ht]{55mm}
    \centerline{
        \epsfig{file=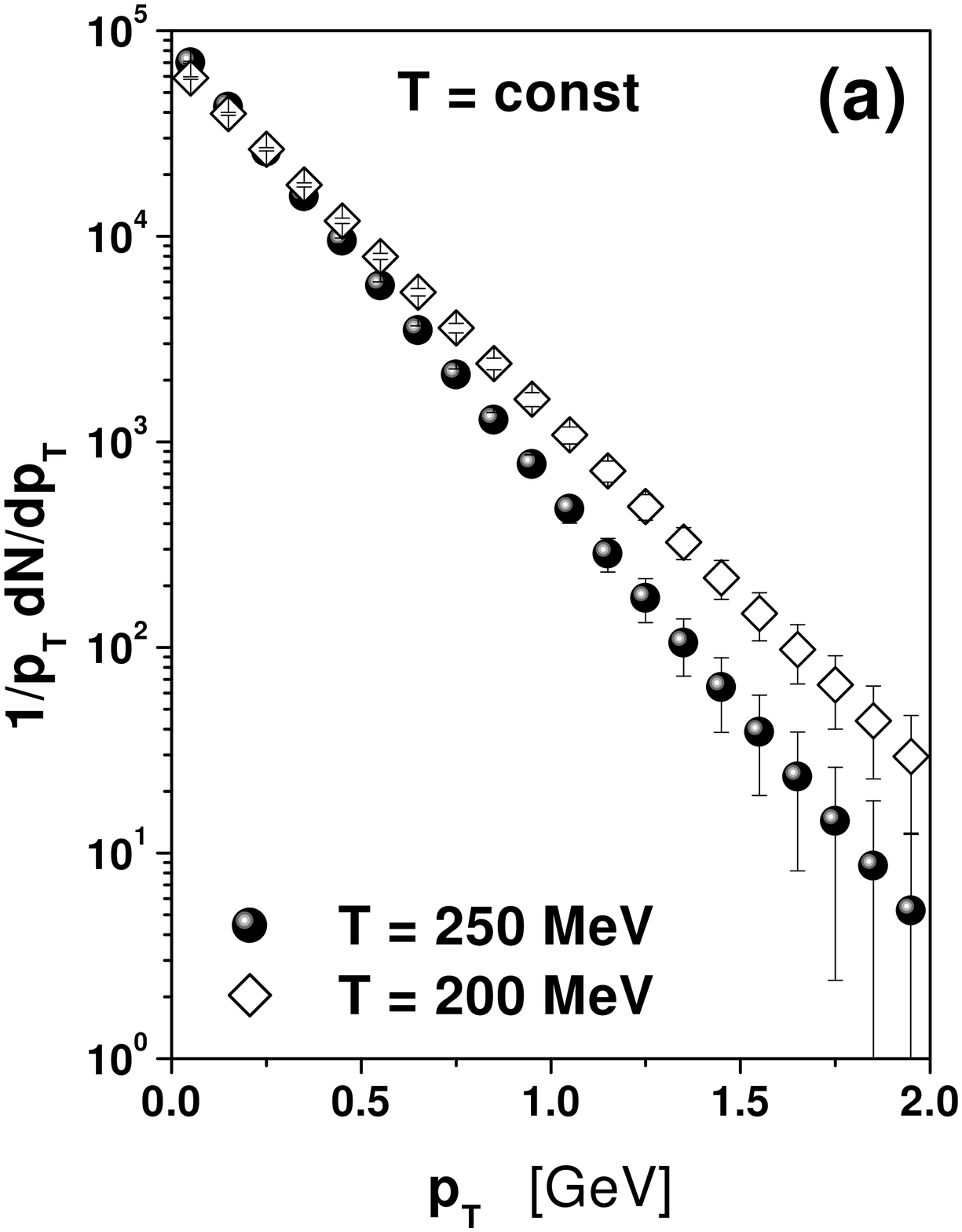, width=45mm}
     }
  \end{minipage}
\hfill
  \begin{minipage}[ht]{55mm}
    \centerline{
       \epsfig{file=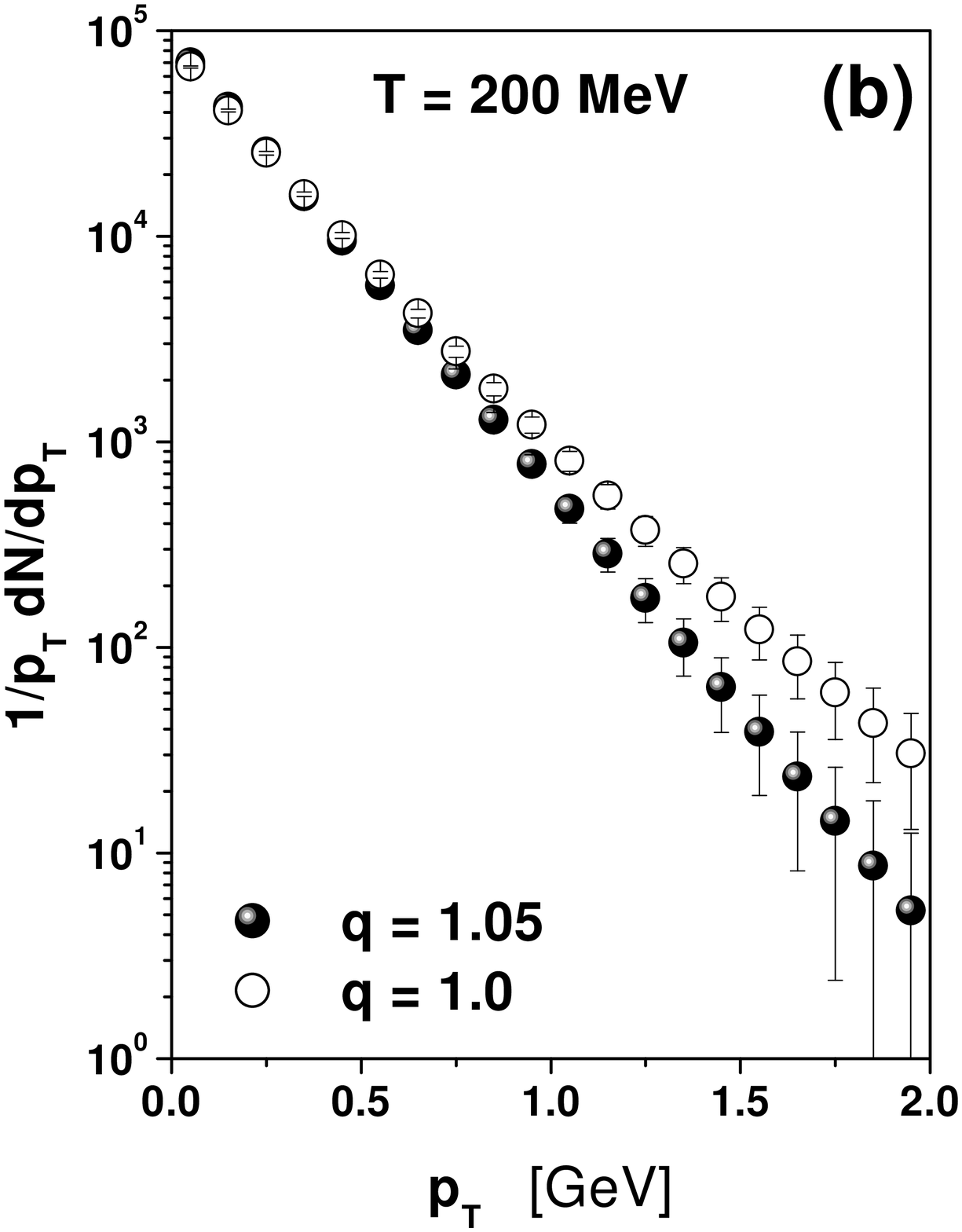, width=45mm}
     }
  \end{minipage}
  \caption{\footnotesize $(a)$ Normal exponential $p_T$ distributions
                     (i.e., $q=1$) for $T = 200$ MeV (black symbols)
                     and $T = 250$ MeV open symbols). $(b)$ Typical
                     event from central $Pb+Pb$ at
                     $E_{beam}=3~A\cdot$TeV (cf. text for other details)
                     for $=200$ MeV for $q=1$ (black symbols) exponential
                     dependence and $q=1.05$ (open symbols).
}
  \label{Figure1}
\vspace{-5mm}
\end{figure}

To observe such fluctuations an event-by-event analysis of data is
needed \cite{OFT}. Two scenarios must be checked: $(a)$ $T$ is
constant in each event but because of different initial conditions it
fluctuates from event to event and $(b)$ $T$ fluctuates in each event
around some mean value $T_0$. Fig. 1 shows typical event obtained in
simulations performed for central $Pb+Pb$ collisions  taking place
for beam energy equal $E_{beam}=3~A\cdot$TeV in which density of
particles in central region (defined by rapidity window $-1.5 <y<
1.5$) is equal to $\frac{dN}{dy} = 6000$ (this is the usual value
given by commonly used event generators \cite{MC}). In case $(a)$ in
each event one expects exponential dependence with $T=T_{event}$ and
possible departure from it would occur only after averaging over
all events. It would reflect fluctuations originating from different
initial conditions for each particular collision. This situation is
illustrated in Fig. 1a where $p_T$ distributions for $ T = 200$ MeV
(black symbols) and $T = 250$ MeV (open symbols) are presented. Such
values of $T$ correspond to typical uncertainties in $T$ expected at
LHC accelerator at CERN. Notice that both curves presented here are
straight lines. In case $(b)$ one should observe departure from the
exponential behaviour already on the single event level and it
should be fully given by $q>1$. It reflects situation when, due to
some intrinsically dynamical reasons, different parts of a given
event can have different temperatures \cite{WW,DENTON}. In Fig. 1b
black symbols represent exponential dependence obtained for $T = 200$
MeV (the same as in Fig. 1a), open symbols show the power-like
dependence as given by (\ref{eq:Pt}) with the same $T$ and with
$q=1.05$ (notice that the corresponding curve bends slightly upward
here). In this typical event we have $\sim 18000$ secondaries, i.e.,
practically the maximal possible number. Notice that points with
highest $p_T$ correspond already to single particles. As one can see,
experimental differentiation between these two scenarios will be very
difficult, although not totally impossible. On the other hand, if
successful it would be very rewarding - as we have stressed before.

One should mention at this point that to the same cathegory of
fluctuating temperature belongs also attempt \cite{BCM} to fit energy
spectra in both the longitudinal and transverse momenta of particles
produced in the $e^+e^-$ annihilation processes at high energies,
novel nonextensive formulation of Hagedorn statistical model of
hadronization process \cite{Beck,Becka} and description of single
particle spectra \cite{UWWF,CERN}.

\subsection{Nonexponential decays}

Another hint for intrinsic fluctuations operating in the physical
system could be the known phenomenon of nonexponential decays
\cite{NEXPDEC}. Spontaneous decays of quantum-mechanical unstable
systems cannot be described by the pure exponential law (neither for
short nor for long times) and survival time probability is
$P(t)\propto t^{- \delta}$ instead of exponential one. It turns out
\cite{NEXPDEC} that by using random matrix approach, such decays can
emerge in a natural way from the possible fluctuations of parameter
$\gamma = 1/\tau$ in the exponential distribution $P(t) =
\exp(-\gamma t)$. Namely, in the case of multichannel decays (with
$\nu$ channels of equal widths involved) one gets fluctuating widths
distributed according to gamma function
\begin{equation}
P_{\nu}(\gamma) = \frac{1}{\Gamma\left(\frac{\nu}{2}\right)}
                    \left(\frac{\nu}{2<\gamma>}\right)
           \left(\frac{\nu \gamma}{2<\gamma>}\right)^{\frac{\nu}{2}-1}
           \exp\left(- \frac{\nu \gamma}{2<\gamma>}\right)
\label{eq:FINALG}
\end{equation}
and strength of their fluctuations is given by relative variance
$\frac{\left\langle (\gamma \, -\,
<\gamma>)^2\right\rangle}{<\gamma>^2}\, = \, \frac{2}{\nu}$, which
decreases with increasing $\nu$.
According to \cite{WW}, it means therefore that, indeed,
\begin{equation}
L_q(t,\tau_0) = \frac{2-q}{\tau_0}\, \left[1\, -\,
(1-q)\frac{t}{\tau_0}\right]^{\frac{1}{1-q}} , \label{eq:Pq}
\end{equation}
with the nonextensivity parameter equal to $q = 1 + \frac{2}{\nu}$.

\section{Summary}

There is steadily growing evidence that some peculiar features
observed in particle and nuclear physics (including cosmic
rays) can be most consistently explained in terms of the suitable
applications of nonextensive statistic of Tsallis. Here we were able
to show only some selected examples, more can be found in
\cite{DENTON,NEXT}. However, there is also some resistance towards
this idea, the best example of which is provided in \cite{GG}. It is
shown there that mean multiplicity of neutral mesons produced in
$p-\bar{p}$ collisions as a function of their mass (in the
range from $m_{\eta} = 0.55$ GeV to  $M_{\Upsilon}=9.5$ GeV) and
the transverse mass $m_T$ spectra of pions (in the range of $m_T
\simeq1\div15$ GeV), both show a remarkable universal behaviour
following over $10$ orders of magnitude the same {\it power law}
function $C\cdot x^{-P}$ (with $x=m$ or $x=m_T$) with $P\simeq 10.1$
and $P\simeq 9.6$, respectively. In this work such a form was just
{\it postulated} whereas it emerges naturally in $q$-statistics
with $q= 1 + 1/P \sim 1.1$ (quite close to results of \cite{BCM}). We
regard it as new, strong imprint of nonextensivity present in
multiparticle production processes (the origin of which remains,
however, to be yet discovered). This interpretation is additionally
supported by the fact that in both cases considered in \cite{GG}
the constant $c$ is the same. Apparently there is no such phenomenon
in $AA$ collisions which has simple and interesting explanation: in
nuclear collisions volume of interaction is much bigger what makes
the heat capacity $C$ also bigger. This in turn, cf. eq.(\ref{eq:C}),
makes $q$ smaller. On should then, indeed, expect that $q_{hadronic}
>> q_{nuclear}$, as observed. 

As closing remark let us point out the alternate way of getting
nonextensive (i.e., with $q \neq 1$) distributions for thermal
models (cf. our remarks in \cite{CR} and the more recent ideas
presented in \cite{Almeida}). Notice that if we allow for the
temperature $T$ to be energy $E$ dependent, i.e., that $T= T(E) = T_0
+ a\cdot (E - \tilde{E})$ (with $a =1/C_V$) then the usual equation
on the probability $P(E)$ that a system $A$ (interacting with the
heat bath $A'$ with temperature $T$) has energy $E$, 
\begin{equation}
d\ln[P(E)] \sim -dE/T \qquad \Longrightarrow \qquad P(E) \sim
\exp\left(- \frac{E}{T}\right)
\end{equation}
becomes 
\begin{equation}
d\ln P(E) \sim \frac{1}{T_0 + a(E-\tilde{E})}dE\quad 
\Longrightarrow\quad
P(E) \sim \left[1 - (1-q)\frac{E-\tilde{E}}{T_0}\right]^{\frac{1}{1-q}}
\end{equation}
with $q=1+a$. This approach could then find its possible application
to studies of fluctuations on event-by-event basis \cite{EBE} (with
all reservations expressed in \cite{EBEW} accounted for).

\section*{Acknowledgments}
GW would like to thank Prof. N.G. Antoniou and all Organizers of X-th
International Workshop on Multiparticle Production, Correlations and
Fluctuations in QCD for financial support and kind hospitality.

\end{document}